\documentclass[APS,twocolumn]{revtex4}
\usepackage{graphics}
\usepackage{amsmath}
\date{\today}

\begin{document}
\title{An optically activated cantilever using photomechanical effects in dye-doped polymer fibers}
\author{Shaoping Bian}
\author{Dirk Robinson}
\author{Mark G. Kuzyk}\email{kuz@wsu.edu}
\affiliation{Department of Physics $\&$ Astronomy, Washington State University, Pullman, Washington 99164-2814}
\begin{abstract}
We report on what we believe is the first demonstration of an optically activated cantilever due to photomechanical effects in a dye-doped polymer optical fiber.  The fiber is observed to bend when light is launched off-axis.  The displacement angle monotonically increases as a function of the distance between the illumination point and the fiber axis, and is consistent with differential light-induced length changes.  The photothermal and photo-reorientation mechanisms, each with its own distinct response time, are proposed to explain the observed time dependence. The measured degree of bending is consistent with a model that we have proposed which includes coupling between photoisomerization and heating.  Most importantly, we have discovered that at high light intensity, a cooperative release of stress results in cis-to-trans isomerization that yields a large and abrupt length change.
\end{abstract}

\maketitle

\section{Introduction}

While the intensity-dependent phase shift of light in a material is often studied, change in the length in a mode of propagation in the material is not usually considered; rather the intensity dependent refractive index and absorption mechanisms are assumed to dominate.  However, there is a long history of observations of the photomechanical effect.  Indeed, in 1966, Merian reported that fabrics doped with azo-dyes contract under light exposure.\cite{merian}  Eisenbach found that a pre-stretched chromophoric poly(ethyl acrylate) network shrinks along the direction of aligned polymer chains when exposed to light.\cite{eisenbach}  More recently, reversible shape changes in solids in the range from 10\% to 400\% and in films up to 20\% that are optically induced by photoisomerization of monodomain nematic elastomers were shown by Finkelmann\cite{finkelmann} and Li et al \cite{li}, respectively.  Camacho-Lopez and coworkers showed that small eleastomer samples that float on water will ``swim" away from regions exposed to light.\cite{camacho-lopez} The strong coupling between orientational order and material strain was described by de Gennes in the early 1970s.\cite{de Gennes}

In this paper, we report the observation that light, which is propagating down a fiber, can be used to induce it to bend, thus functioning as an optically activated cantilever.  Many mechanisms of intensity dependent length change have already been discussed.\cite{welk,zhou} Using the photothermal mechanism,\cite{welker} several devices have been demonstrated including a tunable optical filter,\cite{welker5} an all-optical vibration suppressor\cite{welker2,welker4} and a mesoscale version of such a device that exhibits both mechanical and optical multistability.\cite{welker3} In this paper, we show that differential expansion can be used to make an all-optical cantilever in a dye-doped polymer optical fiber, and we study its mechanisms through modelling and experiment. 

We begin by reviewing some of the more rudimentary observations.  This leads us to the development of our coupled theory of the photomechanical effect - which originates from photothermal heating - and photoisomerization.  This section is followed by a set of experiments that focus on measuring the parameters of the theory, such as the intensity dependence of the time constants and magnitude of bending as a function of pump intensity.  We find that at low intensities, the theory does not behave as expected, because the length decreases rather than increasing with pump intensity.  However, above a threshold intensity, the length change becomes abruptly positive and large -- which is reminiscent of a process in which internal stress builds until a threshold is reached, after which the stress is released.  So, we propose a hypothesis that collective interactions between the molecules are responsible.  We find that our results are consistent with this model; but, we should keep in mind that the model is not meant to be a precise theory.  Rather, its purpose is to deduce the underlying physics of the mechanisms of the response.

\section{Laser induced bending of a fiber}

\begin{figure}
\centering
\includegraphics{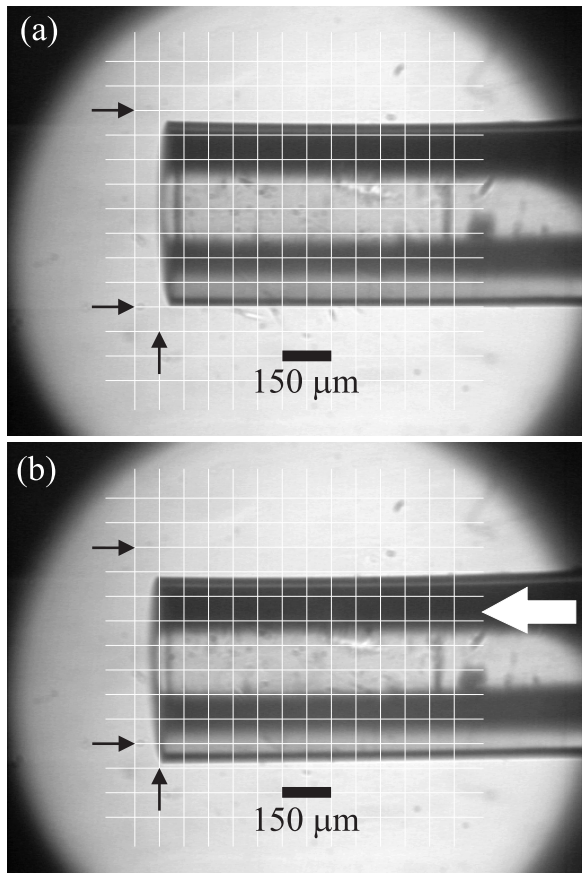}
\caption{A photograph of an MPU (a) without illumination and (b) with illumination.  The three arrows mark reference positions for identifying changes of the fiber position. The white arrow indicates the position of the excitation beam.} \label{fig:bending-photo}
\end{figure}
When a laser beam is launched off-axis into a short fiber made of poly(methyl methacrylate) (PMMA) that is doped with Disperse Red 1 azo-dye (DR1), the fiber is observed to bend. Fig.~\ref{fig:bending-photo}(a) shows a micrograph of a such fiber when no laser beam is launched into it.  The fiber end and sides are aligned to a reticle grid as shown with the three arrows prior to illumination. Fig.~\ref{fig:bending-photo}(b) shows a micrograph of the same part of the fiber when a laser beam is launched in from the right, and the arrow on the right of Fig.~\ref{fig:bending-photo}(b) indicates the launch position.  Relative to the superimposed grid, the fiber's length is observed to increase by about half of a division and bending results in the fiber end moving downwards by almost a full division.  This cantilever is found to move back and forth reproducibly between the two states when the light is repeatedly turned on and off; so the phenomenon is reversible.

\begin{figure}
\centering
\includegraphics{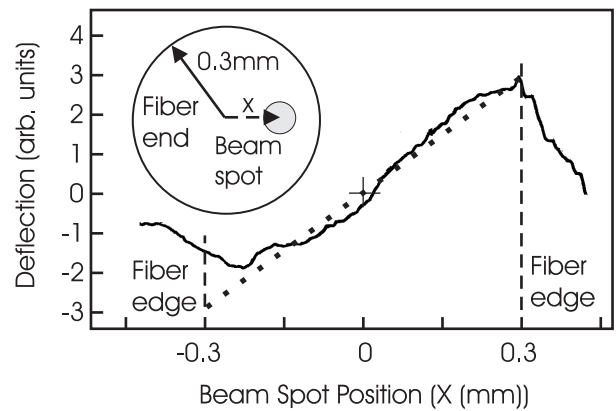}
\caption{The measured degree of fiber bending as a function of the position of the pump laser, measured from the fiber axis.  The inset shows the fiber end and pump beam, approximately to scale.} \label{fig:position-bend}
\end{figure}
The bending is caused by differential expansion between the illuminated and dark portions of the fiber. The illuminated side expands more than the darker side, and the degree of the bending depends on the distance of the pump beam from the fiber axis.  Fig.~\ref{fig:position-bend} shows the measured degree of bending as a function of the pump beam's displacement from the fiber axis.  As expected, the bending angle increases linearly as a function of the axial beam position when the pump beam is fully contained within the fiber.  When the pump is near the fiber edge, some of the light propagates outside the fiber, leading to a decrease in the amount of bending.

\section{Theory}

In this section, we develop a photomechanical theory of heating and photo-reorientation mechanisms.  This calculation is not meant to be complete in that it does not, for example, treat the full three-dimensional problem.  Rather, the goal is to provide a semi-quantitative framework that is applied to understanding the observations.  To this end, several simplifying approximations are made.  These approximations can be understood in terms of the following description.

Consider a polymer optical fiber with an embedded isotropic distribution of one-dimensional chromophores (such as DR1) in the trans state that undergo polarized-laser induced isomerization to a cis state that is bent, and therefore smaller.  The orientational mobility of the cis molecules will be much larger than the trans ones.  The trans molecules, oriented along the polarization direction of a light beam, will have the highest probability of optical absorption.  As such, molecules that are oriented parallel to the laser's polarization axis will be converted to the cis isomer, which subsequently will reorient due to the higher degree of mobility and then decay back to the trans state.  The net result is that molecules oriented parallel to the laser's polarization will be depleted, and converted to an anisotropic distribution, with the net result of a ``hole" in the orientational distribution function.  Consequently, more molecules will be oriented perpendicular to the laser polarization than parallel to it.

We assume that there are two mechanisms of light-molecule interactions that lead to photomechanical effects.  In the photo-reorientation process, photo-isomerization results in a net re-distribution of molecules in a plane that is perpendicular to the light's plane of polarization, which includes a component that is perpendicular to the fiber axis and another component along the fiber axis.  This leads to a stress along the fiber axis resulting in a commensurate change in length and a change in the fiber diameter perpendicular to the beam's polarization, producing an elliptical fiber cross-section.  In the photothermal process, the absorbed energy is turned into heat, which results in thermal expansion.  Since the photo-isomerization process causes an orientational hole, the amount of energy absorbed by the fiber drops over time, so the photothermal process efficiency drops.  In this way, the two mechanisms are coupled.

As a zeroth-order approximation, we assume that photo-isomerization is independent of the temperature for the range of temperature changes expected for photo-heating.  In this case, the photo-reorientation mechanism can be calculated independently of photothermal heating.  The resulting orientational distribution of chromophores due to the photoisomerization process can then be used as a source term in the photothermal calculation, thereby coupling the two mechanisms.  Secondly, we will ignore the effect of the change in fiber diameter and focus only on the change in fiber length.  As such, we will assume that the length change and ellipticity change are decoupled; and, that the ellipticity change does not result in bending.  Indeed, a small part of the cantilever effect that we observe could be attributed to changes in fiber ellipticity; but, if it is proportional to the length change, such an effect could be taken into account as a small correction factor in proportion to the length change.

\subsection{Photo-Reorientation Mechanisms}

A full calculation of the mechanisms would require us to consider a general orientational distribution function, which even for the simple case of the photo-reorientation mechanism alone yields complex results that require numerical integration.  We will therefore simplify the problem by effectively using a two-state model; that is, we assume that a molecule can be oriented only in two orthogonal orientations: perpendicular or parallel to the light beam's polarization.  We will also ignore the cis isomer population and assume that it is small at any instant of time relative to the population of the resulting reoriented trans molecules.

We note that a three-state model could be a better approximation because the molecules that are reoriented away from the polarization direction are equally distributed in the perpendicular plane.  Since on average, we can view such a distribution as half of the molecules being aligned along the fiber axis - resulting in a length change, and the other half being oriented along the fiber diameter - resulting in a diameter change, a three-level model might be more appropriate.  Such an effect can be simply taken into account by attributing a length change to only half the population that is not aligned along the light's polarization direction.  Thus, a fit of the two-level theory to experiment should yield a length change parameter that is double the value that would be given by a three-state model; but should yield the correct dynamics.  So, we proceed in our theoretical development with the understanding that the parameters of our theory will correctly account for the dynamics of our observations; but, that a more rigorous calculation would be required to relate these parameters to real physical properties of the polymer and dopant chromophores.

The approximations we use are as follows:

\begin{enumerate}
\item{The trans molecule interacts with light only if it is oriented along the polarization of the light beam.}
\item{The trans molecule, after interacting with a photon, leads to a trans molecule oriented perpendicular to the photon's polarization.}
\item{When the material is not illuminated, an entropic process causes the system to relax into a steady state equilibrium with equal populations of both orientations.}
\end{enumerate}

We define the following variables:
\begin{itemize}
\item{{\bf $N$} - The fraction of molecules oriented along the polarization of the light beam.  $1-N$ is thus the fraction of molecules perpendicular to the polarization; and, $N=1/2$ is the non-illuminated population.}
\item{{\bf $I$} - The intensity of the light.}
\item{{\bf $\xi$} -  The probability per unit of intensity per unit of time that a trans molecule will absorb light and be converted to the perpendicular orientation.}
\item{{\bf $\beta$} -  The entropic decay rate of an angular hole.}
\end{itemize}

The population, $N$, can thus expressed as
\begin{equation}\label{isomer-rate}
\frac {dN} {dt} = - \xi I  N + \beta (1-2N), 
\end{equation}
where $(1-2N)$ is the difference between the perpendicular and parallel population fraction.  Eq.~(\ref{isomer-rate}) can be integrated to yield,
\begin{equation}\label{isomer-pop}
N = \frac {2 \beta + \xi I  \exp \left[ - \left( 2 \beta + \xi I  \right) t \right]} {2 \left( 2 \beta + \xi I  \right)}, 
\end{equation}
where we have evaluated the integration constant by demanding that $N(t=0) = 1/2$.  Note that we can also express this result in a way that better shows the form of the exponential growth,
\begin{equation}\label{isomer-pop2}
N = \frac {1} {2} \left[1 - \frac { \xi I} {\left( 2 \beta + \xi I  \right)}  \left( 1 - \exp \left[ - \left( 2 \beta + \xi I  \right) t \right] \right) \right] . 
\end{equation}

There are several points about this result that need to be mentioned.  First, at infinite time, the equilibrium population is:
\begin{equation}\label{Neq}
N_{eq} = \frac {\beta} {2 \beta + \xi I } ,
\end{equation}
and the conversion rate of reorientation is intensity dependent and of the form,
\begin{equation}\label{conver-rate}
\delta(I) = 2 \beta + \xi I .
\end{equation}
The limiting forms of Eq.~(\ref{Neq}) are reasonable.  When the entropic decay rate is much larger than the rate of light-induced reorientation ($\beta \gg \xi I $), then $N_{eq} = 1/2$ as expected.  If the decay rate $\beta$ is small compared with the reorientational conversion rate $\xi I $, then all of the parallel population is converted and $N_{eq} = 0$.

When the light source is turned off, if the initial population is $N_{eq}$ and the decay rate is given by Eq.~(\ref{isomer-rate}) with $I = 0$, this yields a population $N$ oriented along the pump beam polarization of 
\begin{eqnarray}\label{pop-decay}
N & = & \frac {1} {2} - \frac {1} {2} \left( 1 - 2 N_{eq} \right) \exp (-2 \beta t) \\ \nonumber & = & \frac {1} {2} - \frac {1}{2}\left[ \frac {\xi I  } {2 \beta + \xi I } \right] \exp (-2 \beta t),
\end{eqnarray}
where we have used Eq.~(\ref{Neq}). Note that Eq.~(\ref{pop-decay}) yields $N(t \rightarrow \infty) = 1/2$.

\subsection{Photothermal Heating Mechanisms}

In this section, we calculate the temperature as a function of time under the assumption that heat is transferred from the light beam to the sample through optical absorption of only the parallel population.  If the light beam is turned off, we assume that the temperature follows Newton's law of cooling.  We define the following parameters:
\begin{itemize}
\item{{\bf $T_0$} - The temperature of the substance surrounding the fiber.}
\item{{\bf $\gamma$} - The cooling rate of the fiber.}
\item{{\bf $\alpha$} -  The temperature increase per trans molecule parallel to the beam's polarization per unit of intensity per unit of time.  Clearly, $\alpha$ depends on the concentration of dopants.}
\end{itemize}

The heating rate of a fiber under {\em uniform illumination} is then given by,
\begin{equation}\label{photo-heat}
\frac {dT} {dt} = -\gamma(T-T_0) + \alpha N I,
\end{equation}
where the first term is the Newton cooling rate and the second term represents energy deposited through optical absorption.  When the light is turned off, the cooling process is described by:
\begin{equation}\label{cool}
\frac {dT} {dt} = -\gamma(T-T_0).
\end{equation}
Clearly, the constant $\gamma$ depends on the fiber geometry and its specific heat.

\bigskip
\noindent \textbf {a). Light-on $\rightarrow$ illumination process}

We begin by considering the heating process.  Substituting Eq.~(\ref{isomer-pop2}) in Eq.~(\ref{photo-heat}) and using Eq.~(\ref{conver-rate}), we get:
\begin{equation}\label{newtIsoHeat}
\frac {dT} {dt} + \gamma(T-T_0) = \frac {\alpha I} {2} \left[ 1 - \frac {(\delta - 2 \beta) } {\delta} [1 - \exp (- \delta t)] \right].
\end{equation}
Integrating Eq.~(\ref{newtIsoHeat}), we get
\begin{equation}\label{newtIsoHeatGeneral}
T - T_0 = A \exp (- \gamma t) + \frac {\alpha\beta I} {\delta \gamma} + \frac { \alpha I} {2 \left( \gamma - \delta  \right) } \frac { \left( \delta - 2 \beta \right) } {\delta} \exp (- \delta t) ,
\end{equation}
where the first term on the righthand side is the homogenous solution with integration constant $A$ and the rest of the expression is the inhomogeneous term.  To find $A$ we demand that at $t=0$ the fiber is in thermal equilibrium with its surroundings so that $T=T_0$.  This yields,
\begin{equation}\label{solveA}
A = - \frac {\alpha\beta I} {\delta \gamma} - \frac { \alpha I} {2 \left( \gamma - \delta  \right) } \frac { \left( \delta - 2 \beta \right) } {\delta} = -\frac{\alpha I(\gamma - 2\beta)}{2\gamma(\gamma -\delta)}.
\end{equation}
Substituting Eq.~(\ref{solveA}) into Eq.~(\ref{newtIsoHeatGeneral}) yields:
\begin{eqnarray}\label{finalHeat}
T - T_0 & = & \frac {\alpha I} {2(\gamma - \delta)} \left[ \left(\frac {\gamma - 2 \beta} {\gamma}\right) [ 1 - \exp (- \gamma t)] \right. \\ \nonumber
& - & \left. \left(\frac {\delta - 2 \beta}{\delta}\right)[1 - \exp (- \delta t)] \right].
\end{eqnarray}

\bigskip
\noindent \textbf {b). Light-off $\rightarrow$ decay process}

Next, we solve for the cooling process, give by Eq.~(\ref{cool}).  This yields,
\begin{equation}\label{finalCool}
T - T_0 = B \exp (- \gamma t) ,
\end{equation}
where $B$ is an integration constant.  The typical experimental sequence consists of turning on the light source, waiting until the fiber reaches equilibrium (i.e. $t \rightarrow \infty$), then turning off the light source.  As such, we choose the temperature difference $T-T_0$ at $t=0$ as the temperature difference from Eq.~(\ref{finalHeat}) at infinite time, or
\begin{equation}\label{HeatEquil}
T - T_0 = \left( \frac {\alpha \beta} {\delta \gamma} \right) I = B,
\end{equation}
so the final cooling result is
\begin{equation}\label{collfini}
T - T_0 = \left( \frac {\alpha \beta} {\delta \gamma} \right) I \exp (- \gamma t) .
\end{equation}

\subsection{Photomechanical Response}

We are now ready to formulate the photomechanical response.  First, we consider the photo-reorientation mechanism.  When a molecule reorients away from the laser's polarization axis, the strain along that axis should decrease and the strain perpendicular to it should increase.   So, the fractional length change, $\Delta L / L$ (where $L$ is fiber length and $\Delta L$ is its change caused by the photomechanical effect) should be proportional to the difference between the fraction of molecules oriented away from the polarization and those parallel to it,
\begin{equation}\label{isomerstrain}
\frac {\Delta L} {L} = b \left[ \left( 1-N \right) - N \right] = b \left( 1-2N \right) ,
\end{equation}
where $b$ is what we call the isomer-mechanical constant, which depends on the properties of the polymer and dopant molecules.  Note that the length changes if the dopant molecules deviate from their isotropic distribution of $N = 1/2$.

In the photothermal mechanism, the length changes due to thermal expansion.  For a coefficient of thermal expansion $\alpha_t$, and assuming additivity between the two mechanisms, the total length change is given by  
\begin{equation}\label{totalstrain}
\frac {\Delta L} {L} =  b \left( 1- 2N \right) + \alpha_t \left(T-T_0 \right) .
\end{equation}

\begin{figure}
\centering
\includegraphics{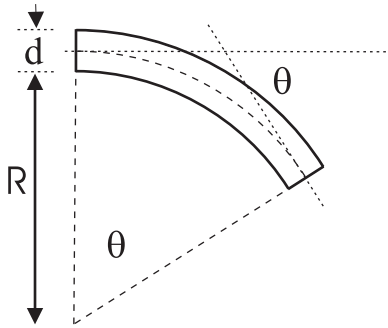}
\caption{A schematic diagram of the relationship between the length change and degree of bending.} \label{fig:bend-calculation}
\end{figure}
Next, we need to determine the relationship between the fractional length change and bending angle of the fiber. As stated above, if the temperature is higher on one side of the fiber, that side will expand, leading to bending away from the hot side.  We assume that half the fiber cross-section is uniformly illuminated while the other half is uniformly dark, so that issues such as continuous temperature gradients are ignored.  Fig.~\ref{fig:bend-calculation} illustrates a bent fiber along with the parameters we use in the following calculation. If the fiber is of length $L$ before bending, we assume that after bending, the length of the fiber along the non-illuminated side remains $L$.  If the length increase of the illuminated side of the fiber is $\Delta L$, we see from Fig.~\ref{fig:bend-calculation} that
\begin{equation}\label{dangle vs dlength}
\theta = \frac {L + \Delta L} {R + d} = \frac{L}{R} \rightarrow  \theta = \frac {L}{d}\frac{\Delta L}{L} = \epsilon\frac {\Delta L}{L} ,
\end{equation}
where $\epsilon = L/d$ is the length-to-diameter ratio of the cantilever, and $R$ is the radius of curvature as shown in Figure \ref{fig:bend-calculation}.

We can calculate the dynamics of photomechanical bending of a fiber by substituting $\Delta L/L$ in Eq.~(\ref{dangle vs dlength}) into Eq.~(\ref{totalstrain}) using the appropriate expression for the number fraction $N$ and the temperature difference $T-T_0$ from the previous section.

It is important to note the effects that we have neglected in the above derivation.  First, we have assumed that half the fiber cross-section is uniformly illuminated and the other half is dark.  Furthermore, we assumed that the temperature increase is proportional to the local intensity, that there are no field gradients, and the the cooling rate depends on the fiber being in contact with a heat batch (i.e. air).

Clearly, the fiber is not illuminated uniformly since the beam has a gaussian profile.  In addition, parts of the beam undoubtedly reflect from the fiber/air interface making the intensity profile even more complicated.  Secondly,  a temperature gradient will form between the illuminated region and the dark side, which complicates the temperature profile and can also cause the beam to refract into the dark region.  These effects together lead our calculations to overestimate the amount of bending and to underestimate the time constant.  Furthermore, since our approximation assumes that the heated side follows its natural length and that the dark side follows suit, this also overestimates the bending angle.  However, our beam is focused so that a region that is smaller than half the cross-section is illuminated, which should result in a larger temperature gradient, thus leading to an underestimate of photo-bending.  Furthermore, we assume that the cylindrical fiber has a rectangular cross-section.  Since the magnitude and time response of the amount of bending that we observe for a polymer fiber agree well with our calculation, we conclude that for the level of accuracy needed to understand the dynamics of the response, our simplified model is sufficient.  Even if a better calculation were available, there are too many unknown parameters for such a model to be of any utility.  As such, we proceed with our simplified model of photothermal heating.

\bigskip
\noindent \textbf {a). Light-on $\rightarrow$ illumination process}

We begin by considering the pump illumination process.  Substituting Equations (\ref{isomer-pop2}) and (\ref{finalHeat}) into (\ref{totalstrain}), and the result into Eq.~(\ref{dangle vs dlength}), the change in angle due to the photoisomerization and photoheating mechanisms, $\theta^r $, is 
\begin{eqnarray}\label{photorise}
\theta^r & = & \frac {\epsilon\alpha_t \alpha I \left( \gamma - 2 \beta \right)} {2 \gamma \left( \gamma - \delta \right)} \left[ 1 - \exp (- \gamma t) \right]\nonumber \\
& + & \frac {\epsilon \left( \delta - 2 \beta \right)  } {\delta} \cdot \left[ b - \frac {\alpha_t \alpha I} {2 \left( \gamma - \delta \right) } \right] \left[ 1 - \exp( - \delta t)\right].
\end{eqnarray}
Note that we have combined terms with the same time constant.  Finally, we substitute Eq.~(\ref{conver-rate}) into Eq.~(\ref{photorise}) to get the result in terms of the material parameters,

\begin{eqnarray}\label{rise}
\theta^r & = & \underbrace{\frac {\epsilon\alpha_t\alpha I \left( \gamma - 2 \beta \right)} {2 \gamma  \left( \gamma - 2 \beta - \xi I \right) } \left[ 1 - \exp ( - \gamma t ) \right]}_{\mbox{Thermal Rise -- Fast}}\nonumber \\ 
& + & \frac { \epsilon\xi I } {2 \beta + \xi I} \cdot \left[ b - \frac {\alpha_t \alpha I} {2 \left( \gamma - 2 \beta - \xi I \right) } \right] \\ \nonumber
& \times & \underbrace{( 1 - \exp \left[ - \left( 2 \beta + \xi I \right) t \right])}_{\mbox{Orientation\: Rise -- Slow}}.
\end{eqnarray}
Each process has associated with it a characteristic time scale. As we will see below, photoisomerization is the slower process, so we label that term the slow process and the heating process term the fast process.

\bigskip
\noindent \textbf {b). Light-off$\rightarrow$ decay process}

Similarly, for the decay process when the pump illumination is off, we substitute Eq.~(\ref{pop-decay}) and (\ref{collfini}) into Eq.~(\ref{totalstrain}) and the result into Eq.~(\ref{dangle vs dlength}).  This yields the angle, $\theta^d$, as a function of time in a dark fiber,

\begin{eqnarray}\label{decay}
\theta^d = \underbrace{\frac{\epsilon\alpha_t\alpha\beta I}{\gamma(2\beta+\xi I)}\exp (-\gamma t)}_{\mbox{Thermal Decay}} + \underbrace{\frac{\epsilon b\xi I}{2\beta+\xi I}\exp (-2\beta t)}_{\mbox{Orientation Decay}}.
\end{eqnarray}
Note that the initial value of Eq.~(\ref{decay}) (at $t = 0$) is equal to that of Eq.~(\ref{rise}) at $t = \infty$.

\section{Experiment}

\begin{figure}
\centering
\includegraphics{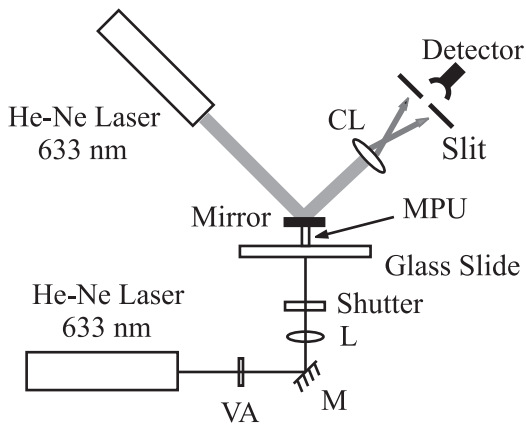}
\caption{A schematic diagram of the experiment used to measure bending of the MPU in response to light excitation.  VA: variable attenuator, M: mirror, L: lens, and CL: cylindrical lens.} \label{fig:experimental-setup}
\end{figure}
To systematically study the photomechanical effects, an experimental setup schematically shown in Fig.~\ref{fig:experimental-setup} is adopted. The fiber used in this experiment is made of PMMA that is uniformly doped with $1\%$ (by weight) DR1 dye. The diameter of the fiber is $600 \, \mu m$ and the length is 2.6$\,$mm, yielding an aspect ratio of $\epsilon = 4.33$.  The fiber's absorption coefficient is $\alpha=0.3 \, mm^{-1}$ at a wavelength of $\lambda = 633 \, nm$, at which the excitation laser operates. We call such fibers mesoscale photomechanical units (MPUs). Details of the fabrication process for making an MPU can be found in the literature.\cite{welker,welker2,welker3} 

The MPU cantilever is polished on both ends, and one end is attached to a glass substrate with transparent cyanoacrylate adhesive. A small piece of a metalized glass cover slip is bonded to the other end and functions as a lightweight mirror.  A beam from a He-Ne laser, called the pump beam, is coupled into the fiber though the substrate. The pump laser beam is focused by a lens and the MPU is located just beyond the focal point, so the spot size of the laser on the input end face of the fiber is estimated to be $\sim 150 \, \mu m$.  An electronic shutter is used to control or modulate the light. A second laser beam, which we call the probe beam, is incident on the MPU's mirror and the reflections from the mirror is directed onto an opaque plate with a narrow slit behind which a photodetector is placed.  A cylindrical lens is inserted between the MPU mirror and detector (see Fig.~\ref{fig:experimental-setup}) to expand the probe beam on the opaque plate thereby increasing the spatial resolution of the slit/detector pair. The pump beam is set off axis near the edge of the fiber for maximum deflection. Also, the pump laser beam is offset horizontally from the fiber axis so that the induced MPU bending occurs in the plane of the table (i.e. in the plane of the page in Fig.~\ref{fig:experimental-setup}) and therefore the movement of the probe laser beam is perpendicular to the slit on the opaque plate. Both pump and probe lasers are He-Ne lasers operating at $\lambda = 633 \, nm$.

\begin{figure}
\centering
\includegraphics{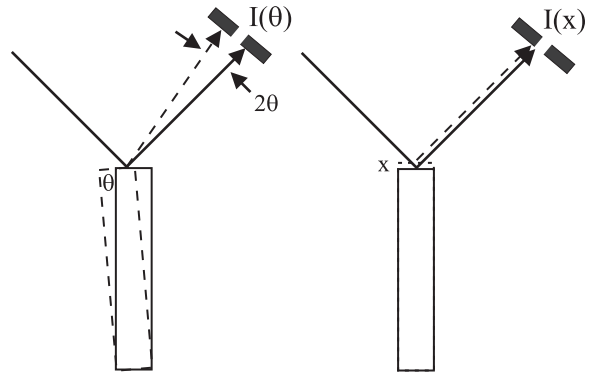}
\caption{A schematic diagram showing how a length change or bending of an MPU affects the measured power at the detector.} \label{fig:bend-shift}
\end{figure}
Fig.~\ref{fig:bend-shift} shows how the change in length and bending affects the beam.  Clearly, the offset of the beam due to a length change is independent of the distance between the MPU's mirror and the detector while bending leads to larger deflections when the distance is made larger.  The change of the detected power due to bending thus depends on this distance and the intensity profile of the laser beam. The greatest sensitivity is achieved when the slit is placed at the point of highest intensity gradient in the beam.  Furthermore, the dynamic range of deflections that can be measured depends on the beam diameter.  Dynamic range is therefore inversely related to the sensitivity. In our experiments, the intensity distribution of the probe laser beam is Gaussian.  We choose the beam size,  focal length of the cylindrical lens, power range of the pump laser, and the slit position such that only the left half (expanded by the cylindrical lens) of the Gaussian beam translates across the slit for the full range of laser-induced MPU deflection. The detector is placed far enough away from the cantilever to increase the sensitivity of detecting bending while minimizing the effect of length change. The output of the detector is fed to a digital oscilloscope in which the experimental data can be saved. In our setup, we have verified that the detected power change due to the MPU's length change is much smaller than that caused by bending of the MPU. Therefore, the power change measured by the detector reflects mainly the effects of bending.  

\begin{figure}
\centering
\includegraphics{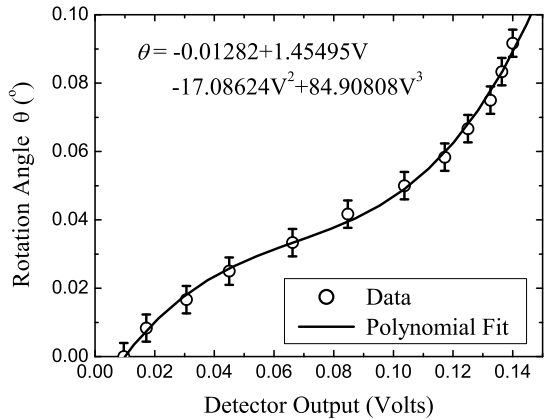}
\caption{The voltage measured at the detector as a function of the mirror deflection angle.  The calibration curve is determined from a polynomial fit.} \label{fig:calibration}
\end{figure}
The apparatus is calibrated to determine the degree of angular deflection per unit of power change of the probe beam.  To do so, the MPU assembly (substrate/MPU/mirror) is placed on a rotation stage with a rotation angle resolution of 0.017$^o$.  The sample is rotated in small, well calibrated angular increments and the power is measured in units of detector voltage.  Fig.~\ref{fig:calibration} shows the data and a polynomial fit.  This calibration polynomial is used to convert the measured probe beam power into a bending angle of the cantilever.  Note that above a deflection angle of $0.1^o$, the displacement of the beam spot at the detector plane is larger than the beam diameter, and therefore beyond the detectable range of the slit/detector combination.  As such, our experiment is limited to deflection angles of less than $0.1^o$.

\section{Results and Discussion}

\begin{figure}
\centering
\includegraphics{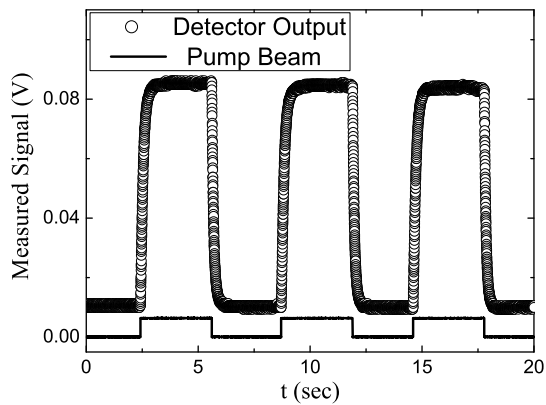}
\caption{Degree of bending of an MPU cantilever as a function of time with pump laser repeatedly turned on and off.} \label{fig:repeat}
\end{figure}
\begin{figure}
\centering
\includegraphics{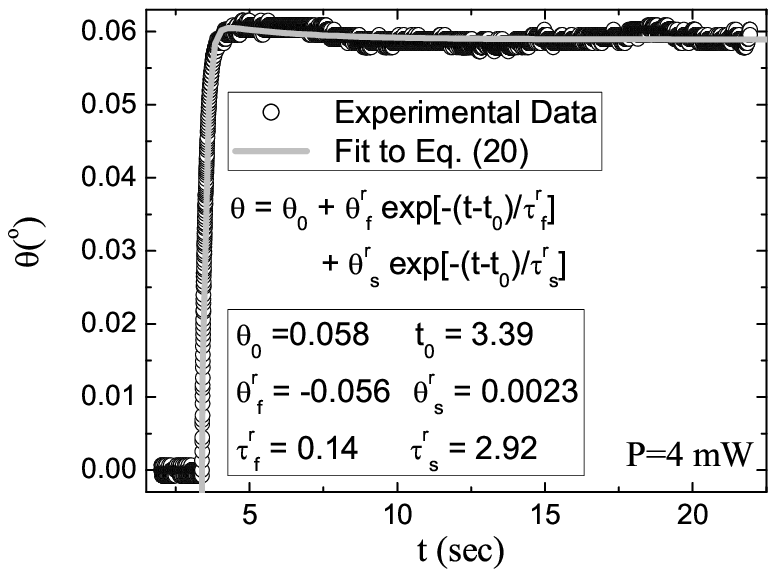}
\caption{The deflection angle of the probe laser (circles) as a function of time after the shutter is opened. The gray curve is a fit to Eq.~(\ref{rise}).} \label{fig:rise}
\end{figure}
\begin{figure}
\centering
\includegraphics{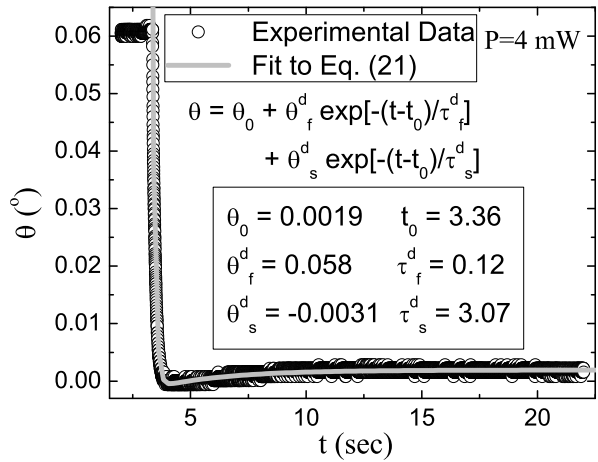}
\caption{The deflection angle of the probe laser (circles) as a function of time after the shutter is closed. The gray curve is a fit to Eq.~(\ref{decay}).} \label{fig:decay}
\end{figure}
The temporal response of the cantilever is obtained by measuring the power at the detector as a function of time after the shutter is opened or closed.  Fig.~\ref{fig:repeat} shows the output of the detector when the shutter is opened and closed repeatedly, which illustrates that the fiber moves reproducibly back and forth. Fig.~\ref{fig:rise} shows a typical plot of the deflection angle (converted from the measured power at the detector by using the calibration polynomial obtained from the fit shown in Fig.~\ref{fig:calibration}) as a function of time after the shutter is opened for a $12.7 \, W/cm^2$ pump intensity. Here we have used an estimated average pump beam diameter of $200 \, \mu m$ inside the MPU. We call the process while the shutter is open a rise process. Fig.~\ref{fig:decay} shows a continuation of Fig.~\ref{fig:rise} when the shutter is closed, which we call a decay process.  The characteristics of these responses are as follows. They can be well fit to Eq.~(\ref{rise}) and (\ref{decay}) respectively, with two different time constants, as shown by the gray curves in Figs.~\ref{fig:rise} and \ref{fig:decay} (all fit parameters are shown as insets).  This indicates that there are two mechanisms that are responsible for the response as modelled by our theory. Furthermore, the two exponential components have opposite signs (there are a few exceptions in the rise data, which we will discuss below). According to Eq.~(\ref{rise}) and (\ref{decay}), it can be concluded that at low intensity,
\begin{equation}
b < 0. \label{b-is-negtive}
\end{equation}
This implies, according to Eq.~(\ref{isomerstrain}), that the bending component caused by the photo-reorientation mechanism is negative, or, photoisomerization yields a decrease in the fiber length. 

\begin{figure}
\centering
\includegraphics{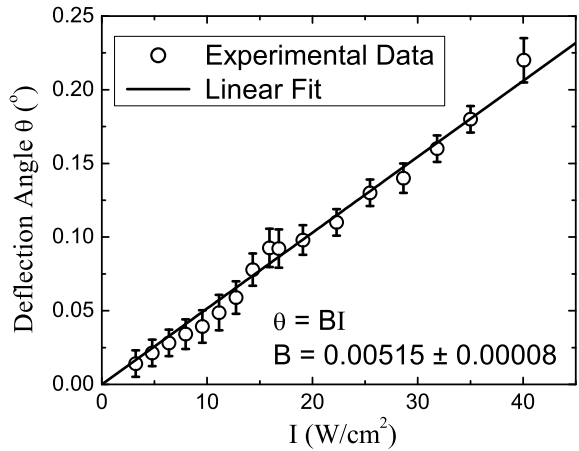}
\caption{The equilibrium deflection angle of the MPU as a function of pump power.} \label{fig:deflection}
\end{figure}
\begin{figure}
\centering
\includegraphics{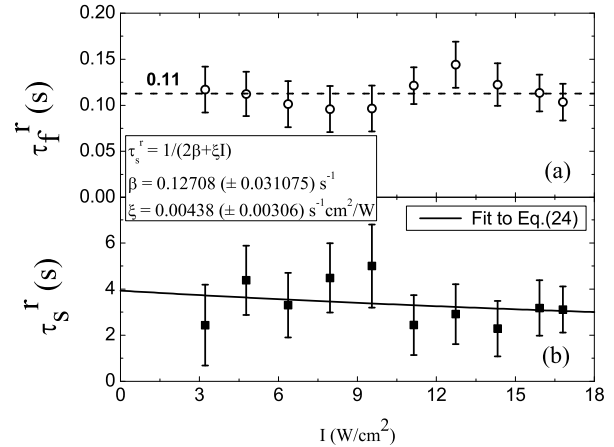}
\caption{Time constants for both the (a) fast and (b) slow process as a function of pump power obtained after the pump is turned on.  Each point in the plot is determined from data fits similar to Fig.~\ref{fig:rise}.} \label{fig:fast-slow-risetime}
\end{figure}
\begin{figure}
\centering
\includegraphics{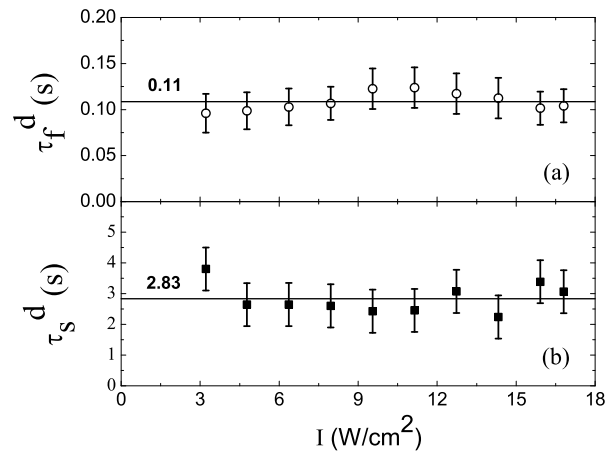}
\caption{Time constants for both the (a) fast and (b) slow process as a function of pump power obtained after the pump pulse is turned off.  Each point in the plot is determined from data fits similar to Fig.~\ref{fig:decay}.} \label{fig:fast-slow-decaytime}
\end{figure}
\begin{figure}
\centering
\includegraphics{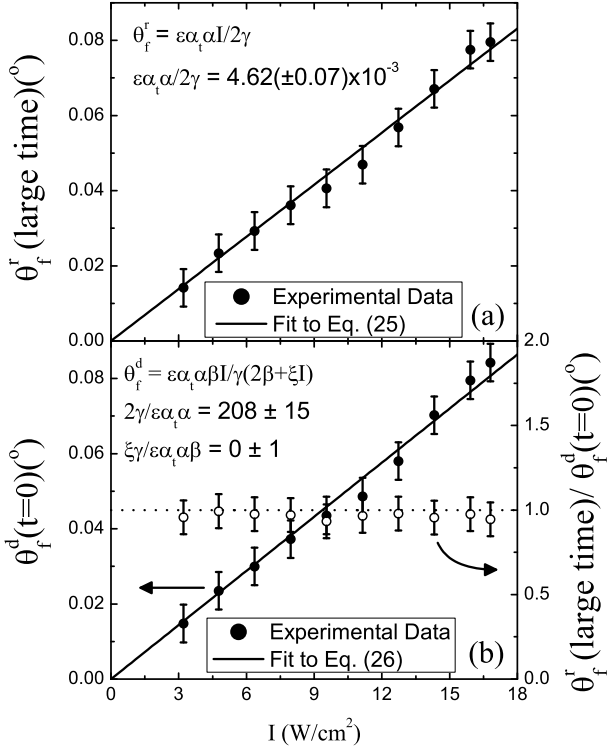}
\caption{The fast components of the amplitude of the deflection angle as a function pump power obtained from (a) rise, and (b) decay data. The open circles indicate the ratio of these two components.} \label{fig:amplitude-fast}
\end{figure}
\begin{figure}
\centering
\includegraphics{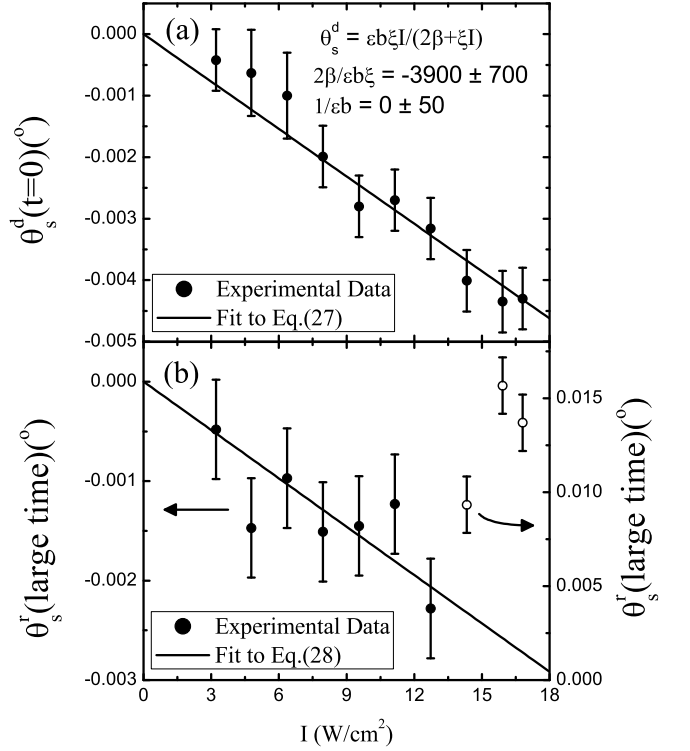}
\caption{The slow component of the amplitude of the deflection angle as a function pump power obtained from (a) rise, and (b) decay data.} \label{fig:amplitude-slow}
\end{figure}

The deflection angle is measured as a function of time after the shutter is opened; and, once equilibrium is reached, after the shutter is closed, at several pump intensities in the range of $3.2 \, W/cm^2$ to $16.9 \, W/cm^2$. Equations (\ref{rise}) and (\ref{decay}) are used to fit these data.  As such, we obtain the power dependence of the physical parameters in our theoretical model, such as the equilibrium steady-state bending angle (Fig.~\ref{fig:deflection}); the rise and decay time constants for the fast and slow mechanisms (Figs.~\ref{fig:fast-slow-risetime} and \ref{fig:fast-slow-decaytime}); and the amplitude of the rise and decay process for the fast and slow mechanisms (Fig.~\ref{fig:amplitude-fast} and \ref{fig:amplitude-slow}).  In Fig.~\ref{fig:deflection}, for intensities above $17W/cm^2$, the deflection angles were measured by observing the reflected spot on a screen and determining the angle trigonometrically.  Any Intensity above $17W/cm^2$ corresponds to a deflection angle greater than $0.1^o$, which is beyond the range of the calibrated slit/detector pair.

\subsection{Time Constants}

According to our theory, the rate, $\gamma$, for heating and cooling are the same and independent of the intensity for the rise and decay processes (see Eq.~(\ref{rise}) and (\ref{decay})). In contrast, the rates of photo-reorientation and its relaxation are different in the rise and decay process, because the former is power dependent (i.e. due to the term 2$\beta + \xi I$ in Eq.~(\ref{rise})) and the latter is power independent (see Eq.~(\ref{decay})). Therefore, the hypothesis that the two time constants are different for the photoisomerization processes is relatively simple to test with intensity-dependent studies of the slow mechanism. 

By comparing the data in Figs.~\ref{fig:fast-slow-risetime} and \ref{fig:fast-slow-decaytime}, we find that within experimental uncertainty, only a weak intensity dependence is observed.  Under some conditions, response times on the order of seconds in the photo-reorientation phenomenon have been reported in the literature.\cite{bian,dumont,todorov,Sekkat92,sekkat01}  However, since the time constant depends on the material composition and intensity, it is not a straightforward matter to determine the time constant that one would expect in our experiments.

To determine which time constant corresponds to which mechanism, we calculate the thermal time constant of the MPU using data from our experiments and PMMA's thermal properties from the literature.\cite{poly} The calculation takes into account the geometry and mass of the fiber and the material's thermal properties (such as the specific heat), and the result indicates a time constant of about a few hundred milliseconds for our experimental conditions, which is consistent with the fast time constants we obtained (shown in Fig.~\ref{fig:fast-slow-risetime}(a) and \ref{fig:fast-slow-decaytime}(a)). Therefore we infer that the fast process originates from photothermal heating, and the slow process is due to photo-reorientation.

An average of the response times in each set of data in Fig.~\ref{fig:fast-slow-risetime} and \ref{fig:fast-slow-decaytime} gives a mean fast response time of $\tau_f^r = 0.113\ (\pm 0.005) \, s$ from the rise process and $\tau_f^d = 0.109\ (\pm 0.003) \, s$ from the decay process; while for the slow process, we get $\tau_s^r = 3.36\ (\pm 0.30) \, s$ from rise data and $\tau_s^d = 2.83\ (\pm 0.15) \, s$ from decay data (these numbers, except $\tau_s^r$, are labelled in Fig.~\ref{fig:fast-slow-risetime} and \ref{fig:fast-slow-decaytime}). The fact that the fast rise and decay time constants are independent of each other and are equal within experimental uncertainty supports our hypothesis that the fast process must be of a photothermal origin.  

In the following section, we extract the various parameters of our model by analyzing the dependence of the data on time and intensity. 

\subsubsection{Heating --- Fast Mechanism}

From the data, an average rise and decay rate of $\gamma = 1/\tau_f = 1/(0.11s) = 9.01\ (\pm 0.16) \, s^{-1}$ is found.

\subsubsection{Photo-Reorientation --- Slow Mechanism\label{POSM1}}

\noindent
\textbf{a). Decay process}
\medskip

The decay rate due to the entropic process of relaxation to the isotropic state is power independent, and is determined from Eq.~(\ref{decay}) to be

\begin{equation}
\beta = \frac{1}{2\tau_s^d} = \frac {1} {2\times 2.83 \, s} = 0.18\ (\pm 0.01) \, s^{-1}.
\label{beta for decay}
\end{equation}

\noindent
\textbf{b). Rise process}
\medskip

The rise time constant originates from the competition between intensity-induced excitation and entropic decay, so is power dependent. Using the relationship between the response time and the intensity from Eq.~(\ref{rise}), the rise time constant is

\begin{equation}
\tau_s^r = \frac {1} {2\beta + \xi I}.
\label{tau for decay}
\end{equation}

Fitting the data in Fig.~\ref{fig:fast-slow-risetime}(b) to this function yields $\beta = 0.13\ (\pm 0.04) \, s^{-1}$ and $\xi = 0.0044\ (\pm 0.0078) \, cm^2W^{-1}s^{-1}$. The value of $\beta$ obtained here from the rise data is 28$\%$ smaller than that obtained from the decay data.  However, they agree with each other within experimental uncertainty.  An average of $\beta = 0.16 \, s^{-1}$ will be used in following calculations. But, we found that scatter in the rise data is relatively larger, and $\chi^2$ of the fit, which is the sum of the squares of deviations of the theoretical curve from experimental points, is 0.95. Therefore, the values obtained from this fit, especially the value of $\xi$, is subject to a large uncertainty. This uncertainty can be further revealed in the following analysis.

\subsection{Amplitude}
\subsubsection{Heating --- Fast Mechanism\label{HFM1}}

\noindent
\textbf {a). Rise process}
\smallskip

In our experimental conditions, we have $\gamma - 2\beta >> \xi I$; and from Eq.~(\ref{rise}) the bending angle of the fast rise process at $t \rightarrow \infty$ is

\begin{equation}\label{fast amplitude for fast rise}
\theta_f^r (t = \infty) = \frac{\epsilon\alpha_t\alpha I}{2\gamma}.
\end{equation}

Fitting the power dependence of the fast component of the bending angle shown in Fig.~\ref{fig:amplitude-fast}(a) to this equation gives the parameter $\alpha_t\alpha = 1.94\ (\pm 0.03)\times 10^{-2} cm^{2}W^{-1}s^{-1}deg$. Here we have used the length-to-diameter ratio $\epsilon = 4.33$ for our cantilever.

\smallskip
\noindent
\textbf{b). Decay process}
\smallskip

According to Eq.~(\ref{decay}), the fast component of the decay of the bending angle at $t = 0$ is

\begin{equation}
\theta_f^d (t = 0) = \frac{\epsilon\alpha_t\alpha\beta I}{\gamma(2\beta + \xi I)}.
\label{fast amplitude for fast decay}
\end{equation}

Fitting the power dependence of the fast amplitude component shown in Fig.~\ref{fig:amplitude-fast}(b) to this equation gives the parameter $\alpha_t\alpha = 2.02\ (\pm 0.01) \times 10^{-2} \, cm^{2}W^{-1}s^{-1}deg$ and a negligible value of $\xi$ ( $< 10^{-11} \, cm^2W^{-1}s^{-1}$).  The value of the parameter $\alpha_t\alpha$ is in good agreement with the one obtained from the rise process. The fact that $\xi$ is negligible is not surprising because the data in Fig.~\ref{fig:amplitude-fast}(b) do not show any evidence of saturation, as predicted by Eq.~(\ref{fast amplitude for fast decay}). Because the $\chi^2$ value of the fit is $8.13 \times 10^{-6}$, it is much better than the fit in Section~\ref{POSM1}b; therefore providing a more accurate value of $\xi$ than we obtained in Section~\ref{POSM1}b.  When $\xi$ is negligible, Eq.~(\ref{fast amplitude for fast decay}) is the same as Eq.~(\ref{fast amplitude for fast rise}), and both amplitudes of rise and decay process follow a linear dependence on pump intensity. The ratio of these two amplitudes calculated from our experimental data is shown in Fig.~\ref{fig:amplitude-fast}(b) (open circles), which demonstrates that the experimental data is consistent with our theory.

\subsubsection{Photo-Reorientation --- Slow Mechanism}

\noindent
\textbf {a). Decay process}
\smallskip

The slow decay amplitude given by Eq.~(\ref{decay}) is 

\begin{equation}\label{amplitude of slow decay}
\theta_s^d (t=0) = \frac{\epsilon b\xi I}{2\beta+\xi I}.
\end{equation}
Fitting Eq.~(\ref{amplitude of slow decay}) to the experimental data in Fig.~\ref{fig:amplitude-slow}(a), we get $b\xi = -1.9\ (\pm 0.3)\times 10^{-5}cm^2W^{-1}s^{-1}deg$, and, again we find that $\xi$ is negligible.  Here we have used an average entropic relaxation rate of $\beta = 0.16 s^{-1}$, as obtained above.  The $\chi^2$ value is $1.58\times 10^{-7}$, so the fit is good.  Note that we kept the term $\xi I$ in the denominator of Eq.~(\ref{amplitude of slow decay}) in our fitting routine so that we could obtain the parameter $\xi$ and compare it with the corresponding one we obtained in Section~\ref{HFM1}b.
    
\smallskip
\noindent \textbf {b). Rise process}
\smallskip

Since $\gamma - 2\beta >> \xi I$, according to Eq.~(\ref{rise}), the amplitude of the photo-reorientation process is

\begin{eqnarray}\label{amplitude of slow rise}
\theta_s^r (t = \infty) & = & \frac { \epsilon\xi I } {2 \beta + \xi I} \cdot \left[ b - \frac {\alpha_t \alpha I} {2 \left( \gamma - 2 \beta \right) } \right].
\end{eqnarray}
Since $b < 0$, $\theta_s^r$ should be negative. The experimental data, as shown in Fig.~\ref{fig:amplitude-slow}(b), are negative at low power, but positive at high power (i.e. the three open circles are of the same sign as the bending angle of the heating process).  A hypothesis that explains the sign reversal will be presented below.  If we fit the negative data of Fig.~\ref{fig:amplitude-slow}(b) only to Eq.~(\ref{amplitude of slow rise}), we obtain $b\xi = -1.2\ (\pm 0.1)\times 10^{-5}cm^2W^{-1}s^{-1} \, deg$, which is about two-thirds of the value we obtained from the decay process. In this fit, the parameter $\xi$, again, is found to be negligible; and so is the second term in brackets in Eq.~(\ref{amplitude of slow rise}) in comparison with $b$. 

\begin{figure}
\centering
\includegraphics{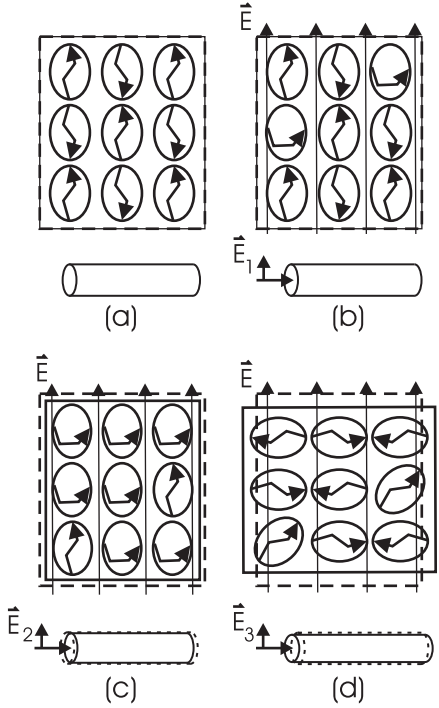}
\caption{Each frame shows a microscopic view of those molecules that are affected most by the light (i.e. we ignore those molecules that are not along the light's polarization) and a macroscopic view of the fiber at a given applied field due to a light beam.  The dashed lines show the original shape of the material or fiber while the solid line shows the instantaneous shape.  Note that $E_1<E_2<E_3$.} \label{fig:SiteDistribute2}
\end{figure}

Our hypothesis for why the amplitude of the slow photoisomerization mechanism starts out negative and abruptly jumps to a large positive value above 13$W/cm^2$ of pump intensity is as follows.  We begin by arguing why a negative angle is observed.  To illustrate the important features of our physical picture, we consider only molecules that are aligned along the direction of the polarization of the light beam since they are the ones that are excited with high probability.  Furthermore, it is well known that the distribution of voids in a polymer vary over a large range of sizes.  For this part of our argument, we only consider those voids that tightly fit around the molecule; and for convenience, represent them as ideal ellipsoids.  Fig.~\ref{fig:SiteDistribute2}(a) shows an illustration of the initial states that are under consideration.  Both the fiber and microscopic view of the molecules are shown.  While the axes of the molecules are randomly oriented so all orientations are equally represented, we only show those molecules in Fig.~\ref{fig:SiteDistribute2} that are along the light's polarization direction.  Also, we show only those voids that are comparable in size to the molecular size (since these are the ones that are responsible for the collective process that we describe below) even though larger voids are present.  Note that we will discuss the role of larger voids at the end of this section.
\begin{figure}
\centering
\includegraphics{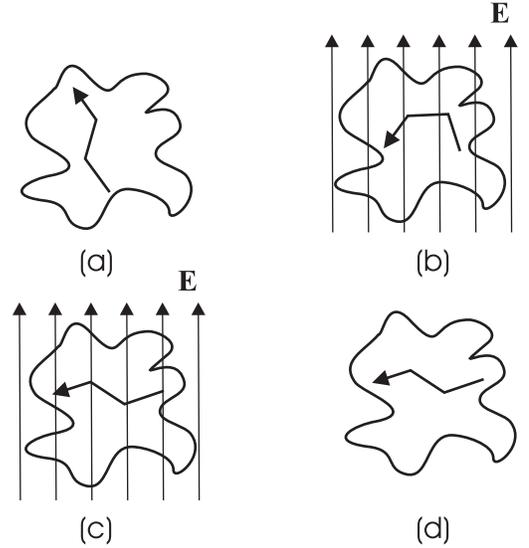}
\caption{The process of trans-cis-trans molecular re-orientation in a void that is large enough to easily accommodate the new orientation.} \label{fig:SiteDistribute}
\end{figure}

A negative angular deflection of the fiber is caused by a decrease in the length of the illuminated side where the polarization is perpendicular to the fiber axis, as shown in Fig.~\ref{fig:SiteDistribute2}(b).  The light therefore causes the molecules that are oriented perpendicular to the fiber (i.e. parallel to the light's polarization) to be excited to the cis state followed by randomization of the orientation of these smaller cis isomers.  Subsequently, if any cis isomer decays back to a trans state that is oriented along the light's polarization, it causes those molecules to be re-excited.  At higher intensity, a significant portion of the trans population becomes depleted as shown in Fig.~\ref{fig:SiteDistribute2}(c).  Since the cis molecules sweep out a smaller volume than the trans ones, we would expect a decrease of void size around the cis molecules, accompanied by a decrease in the polymer's bulk volume, and therefore a decrease in the fiber's length.  While we draw our fibers at high temperature, we have found that there is still a small amount of residual axial stress.\cite{jiang} Shrinkage of a polymer in the direction of stress in response to illumination is commonly observed.\cite{eisenbach}  To summarize, at low intensities (below 13$ \, W/cm^2$), our model assumes that the process follows a sequence of events as shown in Fig.~\ref{fig:SiteDistribute2}(a)-(c).  This process is reversible.

We present the hypothesis that the abrupt transition from negative to positive bend angle arises from a release of internal stress due to the population of cis molecules that are not otherwise conformationally able to relax to the trans state (as shown in Fig.~\ref{fig:SiteDistribute2}(c)).  While the smaller cis molecule may be able to rotate or tunnel into a state with its long axis perpendicular to the cylindrical void, it is difficult for such a cis molecule to decay into a trans state in this orientation.  One can imagine that the cis molecule is being prevented from decaying into the trans state, leading to an internal stress in the polymer whose net direction is on the voids' narrow surface (left to right in Fig.~\ref{fig:SiteDistribute2}(c)).  We thus speculate that if enough molecules accumulate aligned in the same direction in the cis state, they cooperatively force the polymer to expand along the narrow part of the void, allowing them all to collectively decay into the trans state as shown in Fig.~\ref{fig:SiteDistribute2}(d).  So, the aggregate force due to the dyes yields an increase in the fiber's length as shown by the solid lines in the diagram of the fiber in Fig.~\ref{fig:SiteDistribute2}(d).  Though the elongated polymer is in a higher energy state due to increased strain, the overall energy is lower due to the release of stress in the local environment of the interacting monomers and due to the presence of the light beam.  This argument is analogous to ferroelectric domain alignment where the energy of the system is lowered as the surface area of domain walls decreases.  (However, in our model, residual stress remains, as described below.)

If the collective alignment process is elastic in nature (i.e. the polymer remains stressed due to internal elastic forces and conformational changes because the viscosity is negligible on the time scales of the measurement), entropic processes cause fluctuations in alignment that can result in realignment, relieving the residual strain.  The presence of the light, however, prevents enough chromophores from realigning collectively and lowering the energy, so the fiber remains in its elongated state.  Only when the light beam is turned off can entropic processes lead to a collective reorientation accompanied by a decrease of the fiber length.  Since this process is observed to be fully reversible, we argue that the random state of orientation is the favored one, so the presence of the light beam plays an important role in driving the collective reorientational response that leads to a length increase, and without the presence of the light, the random state of orientation is of lowest energy.  Furthermore, while we have focused on the illuminated side of the fiber, clearly, the dark side - which is stressed by the illuminated side - probably also play a role in making a straight fiber the most energetically favored state.

The observation of a small decrease in length at low powers (Fig.~\ref{fig:SiteDistribute2}(c)) and a sharp change to a positive deflection angle at larger power (Fig.~\ref{fig:SiteDistribute2}(d)) is consistent with the picture that at low power, the polymer shrinks because there are not enough cis molecules to collectively overcome the internal stress barrier while at higher powers, the barrier is exceeded.  While illuminated, the length of the fiber is determined by a balance between the stress due to the polymer that acts to decrease the length and the collective force of alignment, which is driven by the light beam.  When the source of illumination is turned off, entropic forces lead to a collective relaxation of chromophore alignment.

For our hypothesis to be plausible, the concentration of DR1 molecules must be large enough to allow cooperative interactions between them.  For the 1\% by weight concentration used in our studies, the mean distance between nearest neighbors is about twice the length of the DR1 molecule.  As such, it is reasonable to expect that molecules near each other can mechanically interact using the polymer as a mediator:  As one void is deformed, the affects are felt by nearby voids as the polymer chains react by steric interactions.

We must reconcile this observation with the literature on the measured refractive index change observed under similar conditions.  The universal observation is that the refractive index change grows monotonically with time and as a function of the intensity.  Furthermore, the measured birefringence shows that molecules are being converted continuously to trans isomers that are perpendicular to the polarization direction of the beam.  This observation appears at first to be inconsistent with our observation of a discontinuity.  We reconcile this apparent contradiction by arguing that the distribution of sites in polymers provides a means for the same microscopic processes to result in different observations for optical and mechanical behavior.  (We note that given the geometry of our experiment, it was not possible to measure the change in absorption spectrum of the illuminated side while operating the cantilever, so the optical properties are deduced from thin-film measurements.)

The distribution of voids in a polymer is known to be large.  As such, we would expect that some of the molecules are loosely held in place with voids whose sizes are approximately equal to the length of the long axis of the trans state while others are held tightly in place by the smaller ellipsoidal voids.  The loosely-bound molecules can easily reorient to any direction in the trans-cis-trans sequence.  Fig.~\ref{fig:SiteDistribute}(a) shows a trans molecule in such a void while Fig.~\ref{fig:SiteDistribute}(b) shows how the cis molecule can freely rotate.  So, we argue that these reoriented dye molecules as shown in Fig.~\ref{fig:SiteDistribute}(c) will yield a negligible stress, which can all be accommodated by small realignment of the chains near the molecule without a global change in the polymer's shape.  So, reorientation can be observed optically, but leaves no mechanical trace.  The tightly bound molecules, however, can only reorient if the local free volume changes shape drastically, which by virtue of the large conformational changes required of the polymer, can only occur if there are global changes in the polymer's shape.  Such global changes require the collective action of many molecules as seen in Fig.~\ref{fig:SiteDistribute2}(d).

\section{Conclusion}

We have demonstrated differential photomechanical effects in dye-doped polymer fibers.  The optically-induced bending of a fiber is governed by two mechanisms with different time response. The analysis suggests the fast one is attributed to a photothermal process whilst the slow one is shown to be related to polymer matrix deformation caused by polarized-laser-induced ordering of the azo-dye molecules.  More importantly, we find what appears to be a new phenomena that leads to photomechanical effects:  At high intensities, internal stress appears to build until there are enough molecules to cause the polymer to reconfigure itself in a way that collectively releases the stress.

Since our work has focused on the fundamental physical basis of the all-optically actuated cantilever, we have intentionally studied systems with low dye concentration because of the resulting simplifications in the modelling.  A higher degree of bending is routinely observed in beams, plates, and films\cite{warner} using isomerization in liquid crystalline materials\cite{finkelmann} and shape-memory polymers.\cite{lendlein}  There are potentially many important applications that may result from miniaturization of devices\cite{yu} such as nano-positioning, optical switching,\cite{athanassiou,hkkim} and beamsteering.  The experimental techniques and theory presented here can be used to study new materials, design devices, and predict their behavior.

\vspace{1em}

{\bf Acknowledgments}
\bigskip

We thank the Washington Technology Center, Sentel Technologies, and the National Science Foundation (ECS-0354736) for supporting this work.

\bibliographystyle{osa}

\end{document}